\documentclass[a4paper,12pt]{article}
\pdfoutput=1
\usepackage{amssymb} 
\usepackage{amsmath}
\usepackage{amsbsy} 
\usepackage{latexsym}
\usepackage{graphicx} 
\usepackage{verbatim}

\newtheorem{theorem}{Theorem} 
\newtheorem{lemma}{Lemma}

\newtheorem{proposition}{Proposition}
\newtheorem{assumption}{Assumption}

\title{Fluid Flow on Vegetated Hillslope}

\author{Stelian Ion$^*$, Dorin Marinescu$^*$, Stefan-Gicu
  Cruceanu\footnote{{Gheorghe Mihoc-Caius Iacob'' Institute
      of Mathematical Statistics and Applied Mathematics,
      Romanian Academy, 050711 Bucharest, Romania, {\tt
        emails: ro\_diff@yahoo.com,
        marinescu.dorin@ismma.ro, gcruceanu@ismma.ro}.}}}

\date{}

\begin{document}
\maketitle

\begin{abstract}
  In this paper, we present a deduction of swallow water
  equations in the presence of vegetation based on spatial
  averaging techniques starting from the general principles
  of conservation of mass and momentum.  For this purpose,
  we worked in the hydrostatic approximation of the pressure
  field and we considered certain hypotheses of kinematic
  and topographical nature and assumptions on the structure
  of the vegetation.  Some elements of differential geometry
  necessary to facilitate the reading of the paper can be
  found in the Appendix.
\end{abstract}

\section{Introduction}
\label{sect_Intro}
The presence of plants on the hill creates a resistance
force to the water flow and influences the process of water
accumulation on the soil surface.  The large diversity of
plants growing on a hill makes the elaboration of an unitary
model of the water flow over a soil covered by vegetation
very difficult.  Here, we present a model based on water
mass and momentum balance equations that takes into account
the presence of certain type of plants.

\begin{figure}
  \centering
  \includegraphics[width=0.7\textwidth]{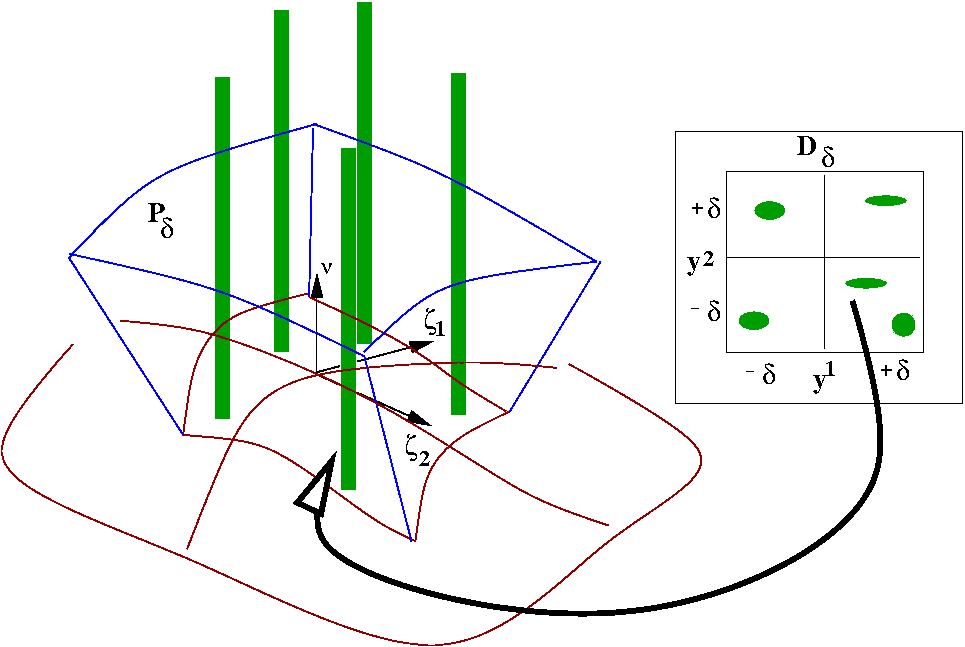}
  \caption{The representative element of the volume
    $P_\delta$ used for mediation. The bottom surface of
    $P_\delta$ has a representative width $\delta$ along two
    orthogonal directions on this surface. The water depth
    $h$ associated to $P_\delta$ is the averaged value of
    the physical water depth $\widetilde{h}$ inside
    $P_\delta$.}
  \label{fig_repr_elem_vol}
\end{figure}

More precisely, the plants form a dense net of rigid
vertical tubes and the water fills the ``voided'' space up
to a level not higher than these plant tubes, see
Figure~\ref{fig_repr_elem_vol}.

The article is structured as follows.  A full hyperbolic PDE
model obtained by averaging the equations for the
conservation of mass and momentum is presented in
Section~\ref{sect_SpAvMod}.  Some closure relations for
these balance equations can be found in
Section~\ref{sect_ClosureRels}, while some mathematical
properties of this model are pointed out in
Section~\ref{section_SWE_models}.  For practical purposes, a
simplified model that preserves the properties of the
general model is also considered in this last section.  The
Appendix is dedicated to some elements of differential
geometry used throughout the paper.

\section{Space Averaging Models}
\label{sect_SpAvMod}
Space averaging is a method to define a unique continuous
model associated to a heterogeneous fluid-solid mechanical
system.  The method is largely used in porous soil media
models \cite{bear, hassan, whitaker}.  For the
fluid-plant physical system, the porous analogy was also
used in \cite{baptist, lowe, nepf}, especially
in the case of submerged vegetation.

At a hydrographic basin scale, there are variations in the
geometrical properties of the terrain (curvature,
orientation, slope) and vegetation density or vegetation
type etc.  Assume there is a map that models the terrain
surface
\begin{equation}
  x^i=b^i(\xi^1,\xi^2),\quad (\xi^1,\xi^2)\in D\subset
  \mathbb{R}^2,\quad i=1,2,3.
	\label{swe_topo_01.eq}
\end{equation}
Denote the tangent vectors to the coordinate curves on this
surface by
\begin{equation}
  \boldsymbol{\varsigma}_a=\partial_a\boldsymbol{b}
  :=\displaystyle\frac{\partial\boldsymbol{b}}{\partial{\xi^a}},\quad a=1,2.
\end{equation}
 
Using this fixed surface, one introduces a new coordinate
$y^3$ along the normal direction $\boldsymbol{\nu }$ to the
surface.  A point in the neighborhood of this surface is
defined in this new system of coordinates
$Y=(\xi^1,\xi^2,y^3)$ by
\begin{equation}
  x^i=b^i(\xi^1,\xi^2)+y^3\nu^i,
  \quad (\xi^1,\xi^2)\in D\subset \mathbb{R}^2,\quad
  y^3\in J\in\mathbb{R},\quad i=1,2,3,
  \label{swe_topo_03.eq}
\end{equation}
where $\boldsymbol{\nu}=(\nu^1,\nu^2,\nu^3)$ represents the
unit normal to the surface.

We introduce the tangent vectors to the coordinate curves
defined by $Y$
\begin{equation}
  \boldsymbol{\zeta}_I:=\partial_I\boldsymbol{x},\quad I=1,2,3.
\end{equation}
One has
\begin{equation}
  \boldsymbol{\zeta}_3=\boldsymbol{\nu},\quad
  \boldsymbol{\zeta}_a=(\delta^b_a-y^3\kappa^b_a)\boldsymbol{\varsigma}_b,
  \quad a=1,2,
  \label{bases_connexion}
\end{equation}
where $\boldsymbol{\kappa} $ is the curvature tensor of the
terrain surface.

In the presence of vegetation on the hill slope, the fluid
occupies the free space between plant bodies and the
mechanical characteristics of the fluid flow are defined
only in the domain occupied by the fluid.

\medskip\noindent
We adopt the following \\
{\bf General convention:} {\it any variable bearing a tilde
  over it designates a micro-local physical quantity, while
  the absence of tilde indicates the corresponding averaged
  quantity. Also, when the micro-local quantity does not
  differ from the corresponding averaged quantity, we
  denote the micro-local quantity without tilde.}

\medskip\noindent Denote by $\Omega_f$ and $\Omega_p$ the
spatial domain occupied by fluid and plants, respectively.
Consider $\widetilde{\psi}$ to be some microscopic quantity
that refers to the fluid.  Let $\boldsymbol{y}=(y^1, y^2)$
be a point in $D$.  One introduces the rectangular domain
\begin{equation}
  D_{\delta}=D_{\delta }(\boldsymbol{y})
  :=[y^1-\delta,y^1+\delta]\times[y^2-\delta,y^2+\delta].
  \label{Dy}
\end{equation} 
Define the spatial averaging volume
\begin{equation*}
  \begin{split}
    P=P(\boldsymbol{y}) =
    \left\{(x^1,x^2,x^3)\left|\right.\right. &
        x^i=b^i(\xi^1,\xi^2)+y^3\nu^i,\\
      & \left. 0<y^3<\bar{h}(\xi^1,\xi^2),\;
        (\xi^1,\xi^2)\in{D}_{\delta}(\boldsymbol{y}),\;
        i=1,2,3\right\}.
  \end{split}
\end{equation*}
Here, $\bar{h}$ is some extension of $\widetilde{h}$ to the
domain $D$, where $\widetilde{h}$ is the function describing the
free water surface outside the domain occupied by plants.

Denote by $P^f$ the fluid domain inside $P$,
\[
P^f:=P\cap\Omega^f.
\]
The boundary of $P^f$ can be partitioned as
\[
\partial
P^f=\Sigma^{fp}\cap\Sigma^{ff}\cap\Sigma^{fa}\cap\Sigma^{fs},
\]
where $\Sigma^{fp}$ is the fluid-plant contact surface
inside $P^f$, $\Sigma^{fa}$ is the free surface of the fluid
inside $P^f$, $\Sigma^{fs}$ is the fluid-soil contact
surface inside $P^f$, and $\Sigma^{ff}$ is the boundary
surface separating the fluid inside and outside $P^f$.

The general form of a balance equation, \cite{muller} is 
\begin{equation}
  \partial_t\int\limits_{P^f}\widetilde{\rho}\,\widetilde{\psi}{\rm d}V+
\int\limits_{\partial P^f}\widetilde{\rho}\,\widetilde{\psi}(\widetilde{\boldsymbol{v}}\cdot\boldsymbol{n}-u_n){\rm d}\sigma=
  \int\limits_{\partial P^f}\widetilde{\boldsymbol{\Phi}}_{\psi}\cdot\boldsymbol{n}{\rm d}\sigma+
  \int\limits_{P^f}\widetilde{\rho}\,\widetilde{\boldsymbol{\phi}}_{\psi}{\rm d}V.
  \label{balance_eq.01}
\end{equation}
Here, the significance of the above quantities are:

- $\widetilde{\rho}$ -- the micro-local mass density of the
fluid;

- $\widetilde{\boldsymbol{v}}$ -- the micro-local velocity of the fluid;

- $\boldsymbol{n}$ -- the exterior unit normal on
$\partial P^f$;

- $\widetilde{\boldsymbol{{\Phi}}}_{\psi}$ -- the micro-local flux density of
$\widetilde{\psi}$;

- $\widetilde{\boldsymbol{\phi}}_{\psi}$ -- the micro-local mass density of
supply $\widetilde{\psi}$;

- $u_n$ -- the normal surface velocity;

- d$V$ -- the volume element;

- d$\sigma $ -- the surface element.

To obtain a mathematical treatable model, one needs to make
some assumptions concerning the complex fluid-plant-soil
system.  The first assumption refers to the plant cover.
\begin{assumption}[Vegetation structure] The plant cover satisfies:\\
  {\rm A1}. The plants are almost normal to the terrain
  surface and they behave like rigid sticks.\\
  {\rm A2}. The water depth is smaller than the height of
  the plants.
\end{assumption}
We remark that A1 is often used in the porous model of the
vegetation and A2 is proper to the overland flow.

The soil-fluid ${\cal I}_{fs}$ and fluid-air ${\cal I}_{fa}$
interfaces can be represented as
\[ {\cal I}_{fs} :=\{\boldsymbol{x}\left|
  x^i=b^i(\xi^1,\xi^2),\quad (\xi^1,\xi^2)\in D^f,\;
  i=1,2,3\right.\}
\]
and
\[ {\cal I}_{fa}:=\{\boldsymbol{x}\left|
  x^i=b^i(\xi^1,\xi^2)+ \widetilde{h}(\xi^1,\xi^2)\delta
  ^i_3,\quad (\xi^1,\xi^2)\in D^f,\; i=1,2,3\right.\},
\]
respectively, where $D^f := \left\{
    (\xi^1,\xi^2)\in D \left| \boldsymbol{b}(\xi^1,\xi^2)
      \in \Omega^f \right.\right\}$.

Define the averaged water depth by
\begin{equation}
  h(y^1,y^2,t):=\displaystyle\frac{1}{\omega_f}\int\limits_{D^f_\delta}
  \widetilde{h}(\xi^1,\xi^2,t)\beta(\xi^1,\xi^2){\rm d}\xi^1{\rm d}\xi^2,
  \label{av.01}
\end{equation}
where $\omega_f$ measures the area of $\Sigma^{fs}$,
\begin{equation}
  \omega_f:=\int\limits_{D^f_\delta}\beta(\xi^1,\xi^2){\rm
    d}\xi^1{\rm d}\xi^2.
  \label{omegaf}
\end{equation}
The volume of the fluid inside the elementary domain $P$ is
given by
\begin{equation}
  {\rm vol}(P^f)=\omega_f h.
	\label{av.02}
\end{equation}

A pure geometrical result which refers to the flux of
$\widetilde{\psi}$ through the boundary $\Sigma^{ff}$ is
formulated as:
\begin{lemma}
  \begin{equation}
    \int\limits_{\Sigma^{ff}}\widetilde{\rho}\,\widetilde{\psi}\,
    \widetilde{\boldsymbol{v}}\cdot \boldsymbol{n}{\rm d}\sigma=
    \partial_a\int\limits_{D^f}\int\limits_0^{\widetilde{h}(\xi^1,\xi^2,t)}
    \widetilde{\rho}\,\widetilde{\psi}\,\widetilde{v}^a
    \Delta{\rm d}y^3\beta(\xi^1,\xi^2){\rm d}\xi^1{\rm d}\xi^2,
  \end{equation}
  \label{lemma_01}
\end{lemma}
where $\Delta=1-y^3K_M+(y^3)^2K_G$, with $K_M$ and $K_G$ the
mean and Gauss curvature respectively, and
$\beta{\rm d}\xi{\rm d}\eta$ is the area element of the
terrain surface.  The quantities $\widetilde{v}^a$, with
$a=1,2$ stand for the contravariant components of the
velocity fields in the local basis
$\{\boldsymbol{\zeta}_I\}_{I=\overline{1,3}}$
\[
\widetilde{\boldsymbol{v}}=\widetilde{v}^a\boldsymbol{\zeta}_a+\widetilde{v}^3\boldsymbol{\nu}.
\]
In Lemma~\ref{lemma_01}, the partial differentiation $\partial_a$
stands for
\[
\partial_a:=\frac{\partial }{y^a}.
\]

\subsection{Averaged mass balance equation}
Although the water density is considered to be a constant
function, we keep it in the mass balance formulation for
emphasizing the physical meaning of the equations.  Define
the averaged water flux by
\begin{equation}
  \rho v^a(\boldsymbol{x},t):=
  \displaystyle\frac{1}{{\rm vol} (P^f)}
  \int\limits_{D^f_\delta }\int\limits_0^{\widetilde{h}(\xi^1,\xi^2,t)}
  \widetilde{\rho}\;\widetilde{v}^a\Delta{\rm d}y^3\beta{\rm d}\xi^1{\rm d}\xi^2.
  \label{wsp-eq.03}
\end{equation}
The mass balance equation results from (\ref{balance_eq.01})
by taking $\widetilde{\psi}=1$, $\widetilde{\boldsymbol{\Phi}}_\psi=0$ and
$\widetilde{\phi}_\psi=0$. Since the plants are treated as
solid bodies and the water does not penetrate the plant
bodies, the water flux through the boundary of the
elementary volume $P^f$ reduces to
\[
\int\limits_{\partial
    P^f}\widetilde{\rho}(\widetilde{\boldsymbol{v}}\cdot\boldsymbol{n}-u_n){\rm
    d}\sigma=\int\limits_{\Sigma^{ff}}\widetilde{\rho}\;
\widetilde{\boldsymbol{v}}\cdot\boldsymbol{n}{\rm
  d}\sigma+
\int\limits_{\Sigma^{fa}}\widetilde{\rho}(\widetilde{\boldsymbol{v}}\cdot\boldsymbol{n}-u_n){\rm
  d}\sigma+
\int\limits_{\Sigma^{fs}}\widetilde{\rho}\;\widetilde{\boldsymbol{v}}\cdot\boldsymbol{n}{\rm
  d}\sigma.
\]
The second integral in the r.h.s. of the above relation
represents the water flux due to the rain which leads to the
water mass gain inside $P^f$.  The third term corresponds to
the water flux due to the infiltration which contributes to
the water loss inside $P^f$.  Using Lemma~\ref{lemma_01} and the
definition of the averaged quantities, one can write the
mass balance:
\begin{equation}
  \displaystyle\frac{\partial}{\partial t}
  \left(\omega_fh\right)+\partial_a\left(\omega_fhv^a\right)
  =\omega r-\omega_fi,
	\label{balance_eq.02}
\end{equation}
with
\begin{equation}
  \int\limits_{\Sigma^{fa}}\widetilde{\rho}(\widetilde{\boldsymbol{v}}
  \cdot\boldsymbol{n}-u_n){\rm d}\sigma=
  -\rho \omega r \quad {\rm and} \quad
  \int\limits_{\Sigma^{fs}}\widetilde{\rho}\;\widetilde{\boldsymbol{v}}\cdot\boldsymbol{n}{\rm
    d}\sigma = \rho \omega_f i
  \label{wsp-eq.08}
\end{equation}
representing the rain and the infiltration rates,
respectively.  Here, as in (\ref{omegaf}), $\omega$ is
defined as
$$
\omega := \int_{D_\delta} \beta(\xi^1,\xi^2) {\rm d} \xi^1
{\rm d} \xi^2.
$$

\subsection{Averaged Momentum Balance Equations}
The momentum balance equation results from
(\ref{balance_eq.01}) with
$\widetilde{\psi}=\widetilde{\boldsymbol{v}}$,
$\widetilde{\boldsymbol{\Phi}}_{\psi}=\widetilde{\boldsymbol{T}}$,
where $\widetilde{\boldsymbol{T}}$ is the stress tensor and
$\widetilde{\phi}_{\psi}=\widetilde{\boldsymbol{f}}$, with
$\widetilde{\boldsymbol{f}}$ denoting the body forces. Here,
we only consider the gravitational force.

In contrast to the planar case, there are some difficulties
in writing component-wise the space averaging balance
momentum equations.  These difficulties appear due to the
point dependence of the local basis.  In the euclidean basis
of $X$, the momentum of the elementary volume $P^f$ is given
by
\[ {\cal
  H}^i(P^f)=\int\limits_{P^f}\widetilde{\rho}\;\widetilde{v}^i{\rm
  d}V.
\]
Using the components of $\widetilde{\boldsymbol{v}}$ in the
basis of $Y$ coordinates, we obtain
\begin{equation} {\cal H}^i(P^f)=\int\limits_{\Sigma^{fs}}
  \int\limits_0^{\widetilde{h}}
  \widetilde{\rho}\,\zeta^i_a\,\widetilde{v}^a\Delta{\rm d}y^3{\rm
    d}\sigma+
  \int\limits_{\Sigma^{fs}}\int\limits_0^{\widetilde{h}}
  \widetilde{\rho}\,\nu^i\,\widetilde{v}^3\Delta{\rm d}y^3{\rm
    d}\sigma,
  \label{Hi}
\end{equation}
which can be rewritten as
\begin{equation} {\cal
    H}^i(P^f)=\varsigma_a^i\int\limits_{\Sigma^{fs}}
  \int\limits_0^{\widetilde{h}}
  \widetilde{\rho}\;\widetilde{v}^a\Delta{\rm d}y^3{\rm d}\sigma+
  \nu^i\int\limits_{\Sigma^{fs}}\int\limits_0^{\widetilde{h}}
  \widetilde{\rho}\;\widetilde{v}^3\Delta{\rm d}y^3{\rm d}\sigma
  +{\cal E}^i_1(\widetilde{\boldsymbol{v}}, P^f).
  \label{HiHi}
\end{equation}
Here and in what follows, we make the following convention:
$\boldsymbol{\varsigma}_a=\boldsymbol{\varsigma}_a(\boldsymbol{y})$,
where $\boldsymbol{y}=(y^1,y^2)$ is the point defining the
domain $D_\delta (\boldsymbol{y})$ from (\ref{Dy}).  When it
appears inside the integral, the unit normal
$\boldsymbol{\nu }$ is a variable quantity depending on the
current point from the domain $D_\delta$, but when it
appears outside the integral, it is the unit normal defined
by the same $\boldsymbol{y}$ as $\boldsymbol{\varsigma}_a$.

The term
$${\cal E}^i_1(\widetilde{\boldsymbol{v}}, P^f):=
\int\limits_{\Sigma^{fs}} \int\limits_0^{\widetilde{h}}
\widetilde{\rho} ( \zeta_a^i-\varsigma_a^i)\widetilde{v}^a
\Delta{\rm d}y^3{\rm d}\sigma
$$
represents an error introduced by neglecting the variation
of the basis $\boldsymbol{\zeta}_I$ along the domain $P^f$.

By averaging, from (\ref{HiHi}) one has
\begin{equation} {\cal H}(P^f)=\rho h\omega_fv^a
  \boldsymbol{\varsigma}_a+ \rho
  h\omega_fv^3\boldsymbol{\nu} +{\cal E}_1(\widetilde{\boldsymbol{v}}, P^f).
  \label{balance_eq.00}
\end{equation}

If one neglects the momentum transfer on the fluid-air and
fluid-soil interfaces, then the flux of the momentum through
the boundary $\partial P^f$ can be reduced to
\[ {\cal F}(\widetilde{\rho}\;\widetilde{\boldsymbol{v}},\partial P^f):=
\int\limits_{\partial P^f}
\widetilde{\rho}\;\widetilde{\boldsymbol{v}}(\widetilde{\boldsymbol{v}}
\cdot\boldsymbol{n}-u_n){\rm d}\sigma=
\int\limits_{\Sigma^{ff}}
\widetilde{\rho}\;\widetilde{\boldsymbol{v}}
(\widetilde{\boldsymbol{v}}\cdot\boldsymbol{n}){\rm
  d}\sigma.
\]
Using Lemma~\ref{lemma_01}, one has
\[ {\cal F}(\widetilde{\rho}\;\widetilde{\boldsymbol{v}},\partial P^f)=
\partial_a\int\limits_{D^f}\int\limits_0^{\widetilde{h}(\xi^1,\xi^2,t)}
\widetilde{\rho}\;\widetilde{\boldsymbol{v}}\;\widetilde{v}^a \Delta{\rm
  d}y^3\beta(\xi^1,\xi^2){\rm d}\xi^1{\rm d}\xi^2,
\]
and then,
\begin{equation}
\begin{split}
  {\cal
    F}(\widetilde{\rho}\;\widetilde{\boldsymbol{v}},\partial P^f)=&\\
  &\partial_a(\rho\omega_fhv^bv^a\boldsymbol{\varsigma}_b)+
  \partial_a(\rho\omega_fhw^{ba}\boldsymbol{\varsigma}_b)+
  \partial_a(\rho\omega_fhv^3v^a\boldsymbol{\nu})+\\
  &{\cal E}_2(\widetilde{v}^2,P^f),
\end{split}
  \label{balance_eq.04-aux} 
\end{equation}
where the fluctuation
$$
\rho w^{ab}:=\frac{1}{\omega_f h}
\int_{\Sigma^f}\int_0^{\widetilde{h}(\xi^1,\xi^2,t)}
\widetilde{\rho}(\widetilde{v}^b-v^b)\widetilde{v}^ay^3
\beta(\xi^1,\xi^2){\rm d} \xi^1{\rm d}\xi^2.
$$

The quantity ${\cal E}_2(\widetilde{v}^2,P^f)$ (as
${\cal E}_1(\widetilde{v},P^f)$ appearing above), represents
the error introduced by approximating the variable local
basis $(\boldsymbol{\zeta}_1(\xi^1,\xi^2,y^3)$,
$\boldsymbol{\zeta}_2(\xi^1,\xi^2,y^3)$,
$\boldsymbol{\nu}(\xi^1,\xi^2,0))$ with the fixed local
basis
$(\boldsymbol{\varsigma}_1,\boldsymbol{\varsigma}_2,\boldsymbol{\nu})$
at $(y^1,y^2,0)$. The quantities ${\cal E}_3$, ${\cal E}_4$
and ${\cal E}_5$ introduced in what follows are errors of
the same nature.

Rel. (\ref{balance_eq.04-aux}) can be rewritten as
\begin{equation}
  \begin{split}
    &{\cal F}(\widetilde{\rho}\;\widetilde{\boldsymbol{v}},\partial P^f)=\\
    =&\partial_a(\rho\omega_fhv^bv^a)\boldsymbol{\varsigma}_b+
    \rho \omega_fhv^bv^a\partial_a\boldsymbol{\varsigma}_b+
    \partial_a(\rho\omega_fhw^{ba})\boldsymbol{\varsigma}_b+
    \rho \omega_fhw^{ba}\partial_a\boldsymbol{\varsigma}_b+\\
    &\partial_a(\rho\omega_fhv^3v^a)\boldsymbol{\nu}+
    \rho \omega_fhv^3v^a\partial_a\boldsymbol{\nu}+
    {\cal E}_2(\widetilde{v}^2,P^f)
    \\
    =&\partial_a(\rho\omega_fhv^bv^a)\boldsymbol{\varsigma}_b+
    \rho \omega_f(hv^bv^a+w^{ba})(\gamma^c_{ab}\boldsymbol{\varsigma_c}+
    \kappa_{ab}\boldsymbol{\nu})+\\
    &\partial_a(\rho\omega_fhw^{ba})\boldsymbol{\varsigma}_b+
    \partial_a(\rho\omega_fhv^3v^a)\boldsymbol{\nu}-
    \rho \omega_fhv^3v^a\kappa^b_a\boldsymbol{\varsigma}_b+
    {\cal E}_2(\widetilde{v}^2,P^f)
    \\
    =&\partial_a(\rho\omega_fh(v^bv^a+w^{ba}))\boldsymbol{\varsigma}_b-
    \rho \omega_fhv^3v^a\kappa^b_a\boldsymbol{\varsigma}_b+
    \rho\omega_f(hv^bv^a+w^{ba})\gamma^c_{ab}\boldsymbol{\varsigma_c}+ \\
    &\rho\omega_f(hv^bv^a+w^{ba})\kappa_{ab}\boldsymbol{\nu}+
    \partial_a(\rho\omega_fhv^3v^a)\boldsymbol{\nu}+
    {\cal E}_2(\widetilde{v}^2,P^f)
    ,
  \end{split}
  \label{balance_eq.04}
\end{equation}
where $\gamma^c_{ab}$ are the Christoffel symbols.

To express the contribution of the stress forces to the
momentum balance
we decompose the stress tensor field $\widetilde{\boldsymbol{T}}$ in two
components: the pressure field $\widetilde{p}$ and the viscous
part of the stress tensor field $\widetilde{\boldsymbol{\tau}}$
\[
\widetilde{\boldsymbol{T}}=-\widetilde{p}\boldsymbol{I}+\widetilde{\boldsymbol{\tau}}.
\]
The flux of the stress vector can now be written as
\[ {\cal F}(\widetilde{\boldsymbol{T}},\partial P_f)= {\cal
  F}(-p\boldsymbol{I},\partial P_f)+ {\cal
  F}(\widetilde{\boldsymbol{\tau}},\partial P_f).
\]
An elementary calculation show that
\begin{equation}
	{\cal F}(-p\boldsymbol{I},\partial
  P_f)=-\int\limits_{D^f}\int\limits_0^{\widetilde{h}(\xi^1,\xi^2,t)}
  \left(\partial_apg^{ab}\boldsymbol{\zeta}_b+\partial_3p\boldsymbol{\nu}\right)\Delta{\rm d}y^3\beta{\rm d}\xi^1{\rm d}\xi^2
	\label{pressure_veg.01}
\end{equation}

The pressure field is determined up to a constant value. If
we subtract the atmospheric pressure from the water
pressure, on the interface fluid-air the pressure must be
zero.  We assume the pressure field to be hydrostatically
distributed.

Let $\boldsymbol{g}=-g\boldsymbol{i}_3$ be the gravitational
force acting on the mass unit. In the local frame of
coordinates related to the free surface of the fluid, this
force has the representation
$$
\boldsymbol{g}=\widetilde{f}^a\boldsymbol{\zeta}_a-\widetilde{f}^3\boldsymbol{\nu}.
$$

\begin{assumption}[Hydrostatic approximation] One assumes that\\
  {\rm A3}. The hydrostatic pressure field has the form
  \[
  \widetilde{p}(\xi^1,\xi^2,y^3)=\widetilde{\rho}\,
  \widetilde{f}^3(\widetilde{h}(\xi^1,\xi^2)-y^3).
  \]
\end{assumption}

We neglect the shear forces on the fluid-air interface, i.e.
\[ {\cal F}(\widetilde{\boldsymbol{\tau}},\Sigma^{fa})=0.
\]
On the fluid-soil interface
the stress vector
$\widetilde{\boldsymbol{t}}:=\widetilde{\boldsymbol{\tau
  }}\cdot\boldsymbol{n}$ can be written as
\[
\widetilde{\boldsymbol{t}} = {\widetilde
  {t}^a\boldsymbol{\zeta}_a}+\widetilde{t}^3\boldsymbol{\nu}.
\]
On the interface soil-water we can write
\begin{equation} {\cal F}(\widetilde{\boldsymbol{\tau}},\Sigma^{fs})=
  \boldsymbol{\varsigma}_a
  \int\limits_{\Sigma^{fs}}\widetilde{t}^a{\rm d}\sigma+
  \boldsymbol{\nu}\int\limits_{\Sigma^{fs}} \widetilde{t}^3{\rm
    d}\sigma+ {\cal E}_3(\widetilde{\boldsymbol{T}},\Sigma^{fs}).
  \label{balance_eq.05-aux}
\end{equation}
Introducing the shear force at the fluid-soil interface
\[
\sigma_s^a=\frac{1}{\rho\omega_f}
\int\limits_{\Sigma^{fs}}\widetilde{t}^a{\rm d}\sigma,
\]
relation (\ref{balance_eq.05-aux}) takes the form
\begin{equation} {\cal F}(\widetilde{\boldsymbol{\tau}},\Sigma^{fs})=
  \boldsymbol{\varsigma}_a \rho \omega_f\sigma_s^a+
  \boldsymbol{\nu} \int\limits_{\Sigma^{fs}} \widetilde{t}^3{\rm
    d}\sigma+ {\cal E}_3(\widetilde{\boldsymbol{T}},\Sigma^{fs}).
  \label{balance_eq.05}
\end{equation}

On the fluid-plant interface 
\begin{equation} {\cal F}(\widetilde{\boldsymbol{\tau}},\Sigma^{fp})=
  \int\limits_{\Sigma^{fp}}\widetilde{\boldsymbol{\tau}}\cdot\boldsymbol{n}{\rm d}\sigma =
  \sum\limits_l \int\limits_{\Sigma_l^{fp}}\widetilde{\boldsymbol{\tau}}
  \cdot\boldsymbol{n}{\rm d}\sigma,
  \label{Ffp}
\end{equation}
where $\Sigma_l^{fp}$ is the fluid-plant
  surface corresponding to the plant $l$. Obviously,
  $\bigcup\limits_l\Sigma_l^{fp}=\Sigma^{fp}$.  Since the
plant stems are supposed to be perpendicular to the ground
surface, (\ref{Ffp}) becomes
\begin{equation} {\cal F}(\widetilde{\boldsymbol{\tau}},\Sigma^{fp})=
  \boldsymbol{\varsigma}_a
  \sum\limits_l\int\limits_{\Sigma^{fp}_l} \widetilde{t}^a{\rm
    d}\sigma +{\cal E}_4(\widetilde{\boldsymbol{\tau}},\Sigma^{fp})
  \label{balance_eq.06-aux}
\end{equation}
and introducing the plant resistance force
\[
\sigma_p^a=\frac{1}{\rho\omega}\sum\limits_l
\int\limits_{\Sigma_l^{fp}} \widetilde{t}^a{\rm d}\sigma,
\]
relation (\ref{balance_eq.06-aux}) becomes
\begin{equation} {\cal F}(\widetilde{\boldsymbol{\tau}},\Sigma^{fp})=
\boldsymbol{\varsigma}_a   \rho\omega\sigma_p^a + 
{\cal E}_4(\widetilde{\boldsymbol{\tau}},\Sigma^{fp}).
  \label{balance_eq.06}
\end{equation}

On the fluid interface of $P^f$,
invoking again Lemma \ref{lemma_01}, the contribution of the
viscous part of the stress tensor on the interface
fluid-fluid takes the form
\[ {\cal F}(\widetilde{\boldsymbol{\tau}},\Sigma^{ff})=
\partial_a\int\limits_{\Sigma^{fs}}\int\limits_0^{\widetilde{h}}
\widetilde{\tau}^{ba}\boldsymbol{\zeta}_b\Delta{\rm d}y^3{\rm
  d}\sigma+
\partial_a\int\limits_{\Sigma^{fs}}\int\limits_0^{\widetilde{h}}
\widetilde{\tau}^{3a}\boldsymbol{\nu}\Delta{\rm d}y^3{\rm
  d}\sigma.
\]

Then, we write the above quantity as,
\begin{equation} 
  {\cal F}(\widetilde{\boldsymbol{\tau}},\Sigma^{ff})=\partial_a(\omega_f
  h\tau^{ba}\boldsymbol{\varsigma}_b)+
  \partial_a (\omega_f h\tau^{3a}\boldsymbol{\nu})+
  {\cal E}_5(\widetilde{\boldsymbol{\tau}}_v,P^f).
  \label{balance_eq.08-aux} 
\end{equation}
Rel. (\ref{balance_eq.08-aux}) implies that
\begin{equation}
  \begin{split}
    {\cal F}(\widetilde{\boldsymbol{\tau}},\Sigma^{ff})=&\\
    =&\partial_a(\omega_f h\tau^{ba})\boldsymbol{\varsigma}_b+
    \omega_f h\tau^{ba}\partial_a\boldsymbol{\varsigma}_b+
    \partial_a (\omega_f h\tau^{3a})\boldsymbol{\nu}+
    \omega_f h\tau^{3a}\partial_a \boldsymbol{\nu}\\
    +&{\cal E}_5(\widetilde{\boldsymbol{\tau}}_v,P^f)
    \\
    =&\partial_a(\omega_f h\tau^{ba})\boldsymbol{\varsigma}_b+
    \omega_fh\tau^{ba}(\gamma^c_{ab}\boldsymbol{\varsigma}_c+
    \kappa_{ab}\boldsymbol{\nu} )+
    \partial_a (\omega_f h\tau^{3a})\boldsymbol{\nu}\\
    -&\omega_f h\tau^{3a}\kappa^b_a\boldsymbol{\varsigma}_b+
    {\cal E}_5(\widetilde{\boldsymbol{\tau}}_v,P^f)
    \\
    =&\partial_a(\omega_f h\tau^{ba})\boldsymbol{\varsigma}_b-
    \omega_f h\tau^{3a}\kappa^b_a\boldsymbol{\varsigma}_b+
    \omega_fh\tau^{ba}\gamma^c_{ab}\boldsymbol{\varsigma}_c+
    \omega_fh\tau^{ba}\kappa_{ab}\boldsymbol{\nu} \\
    +&\partial_a (\omega_f h\tau^{3a})\boldsymbol{\nu}+
    {\cal E}_5(\widetilde{\boldsymbol{\tau}}_v,P^f).
  \end{split}
  \label{balance_eq.08}
\end{equation}

For the supply $\widetilde{\Phi}_\psi$, we only consider the contribution
of the gravitational force. Proceeding by components as in
(\ref{HiHi}), the second term in the r.h.s. of
(\ref{balance_eq.01}) is finally expressed as
\begin{equation}
  \int\limits_{P^f}\widetilde{\rho}\widetilde{\boldsymbol{\phi}}_{\psi}{\rm d}V=
  \int\limits_{D^f}\int\limits_0^{\widetilde{h}(\xi^1,\xi^2,t)}\left(\widetilde{f}^a\boldsymbol{\zeta}_a-\widetilde{f}^3\boldsymbol{\nu}\right)\Delta{\rm d}y^3\beta{\rm d}\xi^1{\rm d}\xi^2
\end{equation}

The relations (\ref{balance_eq.00}, \ref{balance_eq.04},
\ref{pressure_veg.01}, \ref{balance_eq.05},
\ref{balance_eq.06}, \ref{balance_eq.08}) and some order
assumptions are the basis for averaged momentum equations.

The porosity $\theta$ of the plant cover is defined by
  \[
  \theta=\frac{\omega_f}{\omega}.
  \]

  Let $\beta _0 = \beta(y_1,y_2)$, where
  $\boldsymbol{y}=(y^1,y^2)$ is the point defining the
  domain $D_\delta (\boldsymbol{y})$ from (\ref{Dy}).

Let $\epsilon$ be a small parameter.

\begin{assumption}[Kinematical and topographical
  assumptions] Suppose that the physical processes satisfy
  the following properties:\\
  {\rm A4}. {\rm The water depth.} $\widetilde{h}=O(\epsilon)$.\\
  {\rm A5}. {\rm The velocity.} $v^3=O(\epsilon)$.\\
  {\rm A6}. {\rm Geometric assumptions:}\\
  {\rm A6.1}. {\rm Curvature.} The terrain surface curvatures
  and the curvature of the coordinate curves are of order of
  $\epsilon$. This means that locally the surface is almost
  planar.\\ 
  {\rm A6.2}. {\rm Metric tensor.}
  $\beta = \beta _0 +O(\epsilon)$.\\ 
  {\rm A7}. {\rm The averaged dimension }
  $\delta$. $d_p<\!\!<\delta<\!\!<L$ and
  $\delta K_M=O(\epsilon)$.\\
  \label{assum}
\end{assumption}

In what follows, by abuse of notations, we denote $\beta _0$
by $\beta $.

The shallow water type approximation of the averaged
momentum balance for an incompressible fluid results by an
asymptotic analysis.
\begin{theorem}[Averaged momentum equations] Under
  assumptions {\rm A1--A7}, the first order approximation
  for the momentum equations is given by
  \begin{equation}
    \partial_t(h\beta \theta v^a)+\partial_b{\mathfrak F}^{ab}(h,\boldsymbol{v})+
    h\beta\theta\beta^{ab}\partial_aw={\mathfrak G}^a(h,\boldsymbol{v}),\quad a=1,2,
    \label{balance_eq.11}
  \end{equation}
  where
  \[
  w=g(b^3+h\nu^3), \quad (g - \textrm{the
   gravitational acceleration})
  \]

  \[ {\mathfrak F}^{ab}(h,\boldsymbol{v})=h\beta\theta \left(
    v^av^b+w^{ab}-\frac{1}{\rho} \tau^{ab}\right),
  \]

  \[ {\mathfrak G}^a(h,\boldsymbol{v})=\beta\theta\sigma^a_p+
  \beta\theta\sigma^a_s-\gamma^{a}_{bc}\eta^{bc}
  \]
  and
  \[
  \eta^{ac}=h\beta\theta \left(
    v^av^b+w^{ab}-\frac{1}{\rho}\tau^{ab}\right).
  \]

\end{theorem} {\it Sketch of proof}.  Using Assumption
\ref{assum} and relations (\ref{balance_eq.00},
\ref{balance_eq.04}, \ref{balance_eq.05},
\ref{balance_eq.06}, \ref{balance_eq.08}) one can prove that
the terms ${\cal E}_1,\ldots,{\cal E}_5$ are of order
$\epsilon^2$. For $\epsilon<\!\!<1$ these terms as well as
the terms containing the factors $v^3h$, $h\kappa $ or $h^2$
(which are of same order $\epsilon^2$) can be neglected.

The equations (\ref{balance_eq.11}) must be supplemented by
empirical laws concerning the {\it averaged stress tensor}
$\boldsymbol{\tau}$, the {\it averaged vegetation force
  resistance} $\boldsymbol{\sigma}_p$, the {\it averaged
  shear fluid-soil force} $\boldsymbol{\sigma}_s$ and the
{\it averaged fluctuation} $w^{ab}$. These empirical laws
are expressed by functions depending on the averaged
velocity $\boldsymbol{v}$, the averaged water depth $h$ and
a set of parameters $\boldsymbol{\lambda}$ defined by the
characteristics of the plant cover.
\begin{equation}
  \left \{
    \begin{array}{ll}
      \tau^{ab}=\mathfrak{T}^{ab}(\nabla\boldsymbol{v},h,\boldsymbol{\lambda}),\\
      &\\
      \sigma_p^{b}=\mathfrak{S}^{b}_p(\boldsymbol{v},h,\boldsymbol{\lambda}),\\
      &\\
      \sigma_s^{b}=\mathfrak{S}^{b}_s(\boldsymbol{v},h,\boldsymbol{\lambda}),\\
      &\\
      w^{ab}=\mathfrak{W}^{ab}(\boldsymbol{v},h,\boldsymbol{\lambda}).
    \end{array}
  \right.
	\label{empirical_law.00}
\end{equation}

\section{Closure Relations}
\label{sect_ClosureRels}
The averaged models of water flow on a vegetated hillslope
consists of mass balance equation (\ref{balance_eq.02}),
momentum balance equations (\ref{balance_eq.11}) and a set
of empirical relations (\ref{empirical_law.00}).

\medskip
\noindent
{\bf The averaged vegetation force resistance}
\medskip

The most used empirical relations that relate the
vegetation resistance and fluid velocity have the form 
\cite{nepf, baptist}
\begin{equation}
  \sigma^a_p=-\frac{1}{2}C_dmhd|\boldsymbol{v}|v^a,
\end{equation}
where $m$ is the number of stems on the surface $\omega$ and
$d$ is the averaged diameters of the stems. The bed shear
stress
\begin{equation}
  \sigma^a_b=-\frac{g}{C_b^2}|\boldsymbol{v}|v^a,
\end{equation}
$|\boldsymbol{v}|$ being the magnitude of the averaged velocity
{\it i.e.}
\[
|\boldsymbol{v}|^2=\beta_{ab}v^av^b.
\]
One assumes that the viscosity of fluid and the fluctuation
of the velocity field have a small effect as compared with
the bed friction and plant resistance. Therefore the base
model is given by
\begin{equation}
  \begin{split}
\hspace {-2mm}    \displaystyle\frac{\partial}{\partial t}
    \left(h\beta\theta\right)+\partial_a\left(h\beta\theta v^a\right)
    =&\beta( \mathfrak{m}_{r}- \theta\mathfrak{m}_{i}),\\ 
\hspace {-2mm} \displaystyle\frac{\partial }{\partial t} h\theta\beta v^c+
    \displaystyle\frac{\partial }{\partial y^a}
    \theta\beta\,h v^cv^a+
    h\theta\beta\gamma^{c}_{ab}v^av^b
    +h\beta\theta\beta^{ca}\partial_aw=&
    -\beta{\cal K}(h,\theta)|\boldsymbol{v}|v^c.
  \end{split}
	\label{swe_vegm.01}
\end{equation}
The parameter function ${\cal K}(h,\theta)$ is given by
\[ {\cal
  K}(h,\theta)=\frac{1}{2}C_{d}m(\boldsymbol{y})hd+\frac{g\theta}{C_{b}^2}
\]
here $m$ stands for the density number of the stems on
surface area.  In our model, the porosity $\theta$
and the density number $m$ are related by
\[
\theta=1-m\frac{\pi d^2}{4}.
\]
such that one can write
\[ {\cal K}(h,\theta)=\alpha_ph(1-\theta)+\alpha_s\theta,
\]
where the new parameters are given by
\[
\alpha_{p}=\frac{2C_{d}}{\pi
  d},\quad\alpha_{s}=\frac{g}{C_{b}^2}.
\]
 
Note that the system equations modeling the water flow
on an unvegetated hill can be obtained from the model
(\ref{swe_vegm.01}) by simply considering the porosity
$\theta=1$.

\section{SWE models}
\label{section_SWE_models}
The full PDE model for the water flow on vegetated hill is
given by (\ref{swe_vegm.01}).  The system is hyperbolic with
source terms and there is an energy function that is a
conserved quantity in the absence of plants and water-soil
friction.  Also, the model preserves the steady state of the
lake.
\begin{proposition}
  The model {\rm (\ref{swe_vegm.01})} is of hyperbolic type
  with source terms.\\
  {\rm (a)} The conservative form of the system is given by
  \begin{equation}
    \partial_t{\cal H}^{i}(\boldsymbol{y},t,\boldsymbol{u})+
    \partial_a{\cal F}^{ia}(\boldsymbol{y},t,\boldsymbol{u})
    ={\cal P}^{i}(\boldsymbol{y},t,\boldsymbol{u}),
    \label{swe_veg_numerics.01}
  \end{equation}
  where
$$\boldsymbol{u}=\left(
\begin{array}{c}
  h\\
  v^1\\
  v^2
\end{array}
\right) ,\quad {\cal H}(\boldsymbol{y},t,\boldsymbol{u})=
\left(
  \begin{array}{c}
    \beta \theta h\\
    \beta \theta h v^1\\
    \beta \theta h v^2\\
  \end{array}
\right),
$$

$$
{\cal F}(\boldsymbol{y},t,\boldsymbol{u})= \left(
  \begin{array}{cc}
    \beta \theta h v^1&\beta \theta h v^2\\
    \beta \theta(h v^1v^1+
    g\nu^3\beta^{11}h^2/2)\quad&\beta\theta(h v^1v^2+
                                              g\nu^3\beta^{12}h^2/2)\\
    \beta \theta(h v^2v^1+
    g\nu^3\beta^{21}h^2/2)\quad&\beta \theta(h v^2v^2+
                                              g\nu^3\beta^{22}h^2/2)\\
  \end{array}
\right),
$$
and
$$
\begin{array}{l}
{\cal P}(\boldsymbol{y},t,\boldsymbol{u})=\\\\
\left(
  \begin{array}{c}
    \beta(\mathfrak{m}_r-\theta(y)\mathfrak{m}_i)\\
    -\beta\theta h\gamma_{ab}^1v^av^b-
    gh\left[
    \beta \theta \beta^{1a}
    \left(\partial_ax^3+\displaystyle\frac{h}{2}\partial_a\nu^3\right)
    -\displaystyle\frac{h}{2}\nu^3\partial_a\beta\theta\beta^{1a}\right]-\beta{\cal K}|v|v^1\\
    -\beta \theta h\gamma_{ab}^1v^av^b-
    gh\left[
    \beta \theta \beta^{2a}
    \left(\partial_ax^3+\displaystyle\frac{h}{2}\partial_a\nu^3\right)
    -\displaystyle\frac{h}{2}\nu^3\partial_a\beta\theta\beta^{2a}\right]-\beta{\cal K}|v|v^2\\
  \end{array}
\right).
\end{array}
$$
{\rm (b)} For any unitary vector
$\boldsymbol{n}\in\mathbb{R}^3$, the eigenvalue problem
\cite{dafermos}
\begin{equation}
  \left(
    \displaystyle\frac{\partial }{\partial u^i}
    {\cal F}^{ja}n_a-
    \lambda
    \displaystyle\frac{\partial }{\partial u^i}
    {\cal H}^{j}
	\right)r^i=0
	\label{swe_veg_numerics.02}
\end{equation}
has three solutions:
\begin{equation}
  \lambda_{-}=v^an_a-\sqrt{g\nu^3h},\quad
  \lambda_0=v^an_a,\quad
  \lambda_{+}=v^an_a+\sqrt{g\nu^3h}.
  \label{swe_veg_numerics.03}
\end{equation}    
\end{proposition}

\noindent {\it Proof}. In order to prove the existence of
the solution for (\ref{swe_veg_numerics.02}), it is
sufficient to show that
$$
\displaystyle\frac{\partial }{\partial u^i}
{\cal F}^{ja}n_a-\lambda
\displaystyle\frac{\partial }{\partial u^i}
{\cal H}^{j}= \beta\theta\left(
  \begin{array}{ccc}
    \delta&hn_1&hn_2\\
    v^1\delta+g\nu^3h\beta^{1a}n_a&h\delta+hv^1n_1&hv^1n_2\\
    v^2\delta+g\nu^3h\beta^{2a}n_a&hv^2n_1&h\delta+hv^2n_2
  \end{array}
\right),
$$
where $\delta=v^an_a-\lambda$.  The solutions
(\ref{swe_veg_numerics.03}) results then from
straightforward calculations.

\begin{proposition}
  The following properties hold for system
  {\rm (\ref{swe_vegm.01})}:

  (a) it preserves the steady state of a
  lake $$x^3+h\nu^3={\rm constant};$$

  (b) there is a conservative equation for the energy
  \begin{equation}
   \displaystyle
   \frac{\partial }{\partial t}
   h\beta\theta{\cal E}+
   \frac{\partial }{\partial y^a}
   h\beta\theta v^a\left({\cal E}+g\frac{h}{2}\nu^3\right)=
    \beta\left(\left(\mathfrak{M}(-\frac{1}{2}|\boldsymbol{v}|^2+w\right) -{\cal K}|\boldsymbol{v}|^3\right),
  \end{equation}
  where
  $${\cal E}:=\frac{1}{2}|\boldsymbol{v}|^2+g(x^3+\frac{h}{2}\nu^3), \quad \mathfrak{M}=\mathfrak{m}_{r}-\theta\mathfrak{m}_{i}$$

  (c) Bernoulli's law.  At a steady state, in the absence of
  mass source and friction force, the total
  energy
  $${\cal E}^{\rm t}=\frac{1}{2}|\boldsymbol{v}|^2+gx^3+p(\boldsymbol{y},h)$$
  is constant along a current line
  \begin{equation}
    v^a\partial_a{\cal E}^{\rm t}=0.
  \end{equation}
\end{proposition}

\medskip
\noindent
{\bf Simplified model}
\medskip

\noindent
The mathematical model (\ref{swe_vegm.01}) is too
complicated for many practical applications, but it
represents a great start to generate simplified models of
certain realistic problems.  A simplified version of the
full model corresponds to a given soil surface topography
and a given structure of the plant cover.  In what follows,
we introduce a simplified variant of (\ref{swe_vegm.01})
that allows variations in the soil topography and plant
porosity, but for which one must consider small departures
from some constant states.

Assume that the soil surface is represented by
\begin{equation}
  x^1=y^1, \quad x^2=y^2, \quad x^3=z(y^1,y^2)
  \label{swe_vegm_rm.01}
\end{equation}
and the surface is such that the first derivatives of the
function $z(y^1,y^2)$ are small quantities.

\medskip
\noindent
{\bf Assumptions}:\\
{\rm (a)} {\it Geometrical
  assumptions: $$\left|\nabla z\right|^2\approx 0, \quad \nabla^2 z\approx 0.$$}

One these grounds, equations (\ref{swe_vegm.01}) can be
approximated as
\begin{equation}
  \begin{split}
    \displaystyle\frac{\partial}{\partial t}\theta h+\partial_a\left(\theta h v^a\right)
    &=\mathfrak{M},\\
    \frac{\partial }{\partial t}
    \theta hv_a+\partial_b \theta hv_av^b+\theta h\partial_aw&=-{\cal K}(h,\theta)|\boldsymbol{v}|v_a,
  \end{split}
  \label{swe_vegm_rm.02}
\end{equation}
where
\begin{equation}
  {\cal K}(h,\theta)=\alpha_ph(1-\theta)+\theta\alpha_s,\quad
  \mathfrak{M}=\mathfrak{m}_{r}- \mathfrak{m}_{i}\theta,\quad
  w=g(z(y^1,y^2)+h).
  \label{swe_vegm_rm.03} 
\end{equation}

The simplified model (\ref{swe_vegm_rm.02}) preserves the
main properties of the full model.
\begin{proposition}
  The reduced model {\rm (\ref{swe_vegm_rm.02})} of equations
  for the water flow on vegetated hill is of hyperbolic type
  with source terms.\\
  {\rm (a)} The conservative form of the system is given by
  \begin{equation}
    \begin{split}
      \displaystyle\frac{\partial}{\partial t}\theta h+\partial_a\left(\theta h v^a\right)
      &=\mathfrak{M},\\
      \displaystyle\frac{\partial }{\partial t}
      \theta hv_a+\partial_b\left(\theta hv_av^b+\delta_a^b \theta g\frac{h^2}{2}\right)&= 
                                                                                                      -hg\partial_az-g\frac{h^2}{2}\partial_a\theta
                                                                                                      -{\cal K}(h,\theta)|\boldsymbol{v}|v_a.
    \end{split}
    \label{swe_veg_numerics.06}
  \end{equation}
  {\rm (b)} For any unitary vector
  $\boldsymbol{n}\in\mathbb{R}^3$, the solutions of the
  eigenvalue problem are given by
  \begin{equation}
    \lambda_{-}=v^an_a-\sqrt{g h},\quad
    \lambda_0=v^an_a,\quad
    \lambda_{+}=v^an_a+\sqrt{g h}.
    \label{swe_veg_numerics.07}
  \end{equation}
\end{proposition}

\begin{proposition}
  The system {\rm (\ref{swe_vegm_rm.02})} has the following
  properties:

  {\rm (a)} it preserves the steady state of a
  lake $$x^3+h={\rm constant};$$

  {\rm (b)} there is a conservative form of the equation for
  the energy dissipation
  \begin{equation}
    \displaystyle\frac{\partial }{\partial t}
    \theta h{\cal E}+
    \frac{\partial }{\partial y^a}
    \theta h v^a\left({\cal E}+g\frac{h}{2}\right)=
    \left(\left(\mathfrak{M}(-\frac{1}{2}|\boldsymbol{v}|^2+w\right) -{\cal K}|\boldsymbol{v}|^3\right),
    \label{swe_veg_numerics.08}
  \end{equation}
  where $${\cal E}:=\frac{1}{2}|\boldsymbol{v}|^2+g\left(x^3+\frac{h}{2}\right);$$

  {\rm (c)} Bernoulli's law.  At a steady state, in the
  absence of mass source and friction force, the total
  energy $${\cal E}^{\rm t}=\frac{1}{2}|\boldsymbol{v}|^2+g x^3+p(\boldsymbol{y},h)$$
  is constant along of a current line
  \begin{equation}
    v^a\partial_a{\cal E}^{\rm t}=0.
  \end{equation}
\end{proposition}

The presence of the plants and the existence of the
frictional interaction between water and soil induce an
energetic loss.  To put in evidence such phenomenon, let us
consider a domain $\Omega$ and $\boldsymbol{n}$ the unitary
normal to $\partial \Omega$ outward orientated.  One
assumes that $\partial \Omega$ consists of an
impermeable portion and an exit portion
$\partial \Omega=\Gamma_1\cup\Gamma_2$,
$\boldsymbol{n}\cdot\boldsymbol{v}=0$ on $\Gamma_1$ and
$\boldsymbol{n}\cdot\boldsymbol{v}>0$ on $\Gamma_2$.  One of
the two portions can be a void set.

\begin{proposition}[Energy disipation] Assume that there is
  no mass production. Then the energy of $\Omega$ is a
  decreasing function with respect to time
  \begin{equation}
    \partial_t\int\limits_{\Omega}h\beta\theta{\cal E}{\rm d}x<0.
    \label{swe_veg_numerics.09}
  \end{equation}
\end{proposition}

To prove the assertion, one integrates the energy
dissipation equation (\ref{swe_veg_numerics.08})
$$
\partial_t\int\limits_{\Omega}h\beta\theta{\cal E}{\rm
  d}x+\int\limits_{\partial\Omega}h\beta\theta\boldsymbol{v}\cdot\boldsymbol{n}{\cal
  E}^t{\rm d}s= -\int\limits_{\Omega}{\cal K}|\boldsymbol{v}|^3{\rm d}x
$$  
and observes that the second integral from the l.h.s. is a
positive quantity.

\section{Conclusion}
\label{sect_Concl}
Using techniques similar to the ones used for the standard
SWE, we presented here a deduction of the SWE with
vegetation.  Mathematical and relevant physical properties
from the standard equations can be found for the new model.
For practical applications, a simplified model is also
constructed and presented in this paper.  This model
successfully preserves the main properties of the full
model.

\appendix
\section{Basics of differential geometry in $\mathbb{E}^3$}

\subsection{Curvilinear coordinate}
Let $O\boldsymbol{x}$ be a Cartesian coordinate system in
the reference Euclidean space $\mathbb{E}^3$.  Let
$\{y^I\}_{I=\overline{1,3}}$ be another coordinate system
and let
\begin{equation}
  x^i=x^i(y^1,y^2,y^3), \quad \boldsymbol{y}\in D
  \label{bdiff_01.eq}
\end{equation} 
be the transformation rule\index{coordinates transform}.  By
coordinate line, one understands the curves generated by the
variation of a single variable $y^I$, while the rest are
kept constants.  The tangent vectors at the coordinate lines
are defined by
\begin{equation}
  \boldsymbol{e}_I=\partial_I\boldsymbol{x}.
  \label{bdiff_02.eq}
\end{equation}
The set of vectors
$\{\boldsymbol{e}_I\}_{I= \overline{1,3}}$ give rise to a
new base of tensor fields.  For the vectors and tensors of
rank $2$, one writes
\begin{equation*}
  \boldsymbol{v}=v^I\boldsymbol{e}_{I},\quad \boldsymbol{\mathfrak{t}}=
  \mathfrak{t}^{IJ}\boldsymbol{e}_{I}\boldsymbol{e}_{J}.
\end{equation*}  
In the new coordinate system, the components of the metric
tensor $\boldsymbol{g}$ are given by
\begin{equation}
  g_{IJ}=\delta_{ij}e^i_Ie^j_J
  \label{bdiff_03.eq}
\end{equation} 
and
\begin{equation}
  g^{IJ}=\delta^{ij}h^I_ih^J_j,
  \label{bdiff_04.eq}
\end{equation}
where
\begin{equation}
  h^I_j=\partial_jy^I.
  \label{bdiff_05.eq}
\end{equation}
One has
\begin{equation}
  e^j_Ih^I_i=\delta_i^j,\quad e^j_Ih^J_j=\delta_I^J
  \label{bdiff_051.eq}
\end{equation}
and then
\begin{equation*}
  g^{IK}g_{KJ}=\delta^I_J.
\end{equation*}
The volume element is
\begin{equation}
  J=\varepsilon_{ijk}e^i_1e^j_2e^k_3,
 \label{bdiff_06.eq}
\end{equation}
with $\varepsilon_{ijk}$ representing the Levi-Civita
symbol.  From (\ref{bdiff_06.eq}) and (\ref{bdiff_03.eq}),
one obtains
\begin{equation}
  {\rm det}\,\boldsymbol{g}=J^2,
  \label{bdiff_07.eq}
\end{equation}
where $\boldsymbol{g}$ is the matrix with the elements $g_{IJ}$.

The variation of the basis $\{\boldsymbol{e}_I\}_I$ with
respect to the $y$ coordinate is stored inside Christoffel's
symbols $\Gamma$
\begin{equation}
  \partial_I\boldsymbol{e}_J=\Gamma^L_{IJ}\boldsymbol{e}_L.
  \label{bdiff_08.eq}
\end{equation} 
Alternatively, one can calculate the $\Gamma$ coefficients
by
\begin{equation}
  \begin{split}
    \Gamma^L_{IJ}&=h^L_i\partial_Je^i_I,\\
    \Gamma^L_{IJ}&=-e^i_Ie^j_J\partial_ih^L_j,\\
    \Gamma^L_{IJ}&=\displaystyle\frac{1}{2}g^{LK}
                    \left(\partial_Ig_{KJ}+\partial_Jg_{KI}-\partial_Kg_{IJ}\right).
  \end{split}
  \label{bdiff_09.eq}
\end{equation}
The first relation here results from the definition
(\ref{bdiff_08.eq}) and (\ref{bdiff_051.eq}), the second
relation results from the first one, and the last relation
results from (\ref{bdiff_08.eq}) and (\ref{bdiff_03.eq}).
Define now the covariant derivative of a vector by
\begin{equation}
  v^I_{;L}=\partial_Lv^I+v^K\Gamma^I_{LK}
  \label{bdiff_10.eq}
\end{equation}
and the covariant derivative of tensor by
\begin{equation}
  t^{IJ}_{;L}=\partial_Lt^{IJ}+t^{KJ}\Gamma^I_{LK}++t^{IK}\Gamma^J_{LK}.
  \label{bdiff_11.eq}
\end{equation}
An elementary way to introduce the covariant
derivative\index{covariant derivative} is to estimate the
difference of vector fields between two neighbor points
\[
\begin{split}
\boldsymbol{v}(\boldsymbol{y}+\triangle\boldsymbol{y})-\boldsymbol{v}(\boldsymbol{y})=
&v^I(\boldsymbol{y}+\triangle\boldsymbol{y})
\boldsymbol{e}_I(\boldsymbol{y})(\boldsymbol{y}+\triangle\boldsymbol{y})
-v^I\boldsymbol{e}_I(\boldsymbol{y})\\
=&\left(\partial_Lv^I(\boldsymbol{y})+
v^K(\boldsymbol{y})\Gamma^I_{LK}(\boldsymbol{y})\right)\boldsymbol{e}_I(\boldsymbol{y})
\triangle y^L + O(\triangle \boldsymbol{y}^2).
\end{split}
\]

\subsection{Basic notions of differential geometry on a
  surface in $\mathbb{E}^3$}
For completeness, we present here the essential facts about
the differential geometry of the surface in the euclidean
space $E^3$; as a reference, one can consult the classical
books \cite{eisenhart}.  Let $O\boldsymbol{x}$ be a
Cartesian coordinate system in the reference Euclidean space
$\mathbb{E}^3$.  Let ${\cal S}$ be a surface in $E^3$ and
let
\begin{equation}
  x^i=b^i(y^1,y^2),\quad (y^1,y^2)\in {D}\in \mathbb{R}^2
  \label{swe_topo_01_00.eq}
\end{equation}
be a parameterization of ${\cal S}$.  One defines the tangent
vectors to the surface by
\begin{equation}
  \tau^i_a=\frac{\partial b^i}{\partial y^a}
  \label{swe_topo_01_01.eq}
\end{equation}
and the oriented normal direction to the surface by
\begin{equation}
  {\cal N}_i=\varepsilon_{j\,k\,i}\tau^j_1\tau^k_2.
  \label{swe_topo_02_02.eq}
\end{equation}
The unitary normal $\boldsymbol{\nu}$ to the surface is
given by
\begin{equation}
  \nu_i=\frac{{\cal N}_i}{||\boldsymbol{\cal N}||}.
  \label{swe_topo_02_03.eq}
\end{equation}

\noindent {\bf Metric tensor 
  $\boldsymbol{\beta}$ of the
  surface}. The covariant components of $\boldsymbol{\beta }$
are given by
\begin{equation}
  \beta_{ab}=\delta_{ij}\tau^i_a\tau^j_b
  \label{swe_topo_02_03_aa.eq}
\end{equation} 
and the contravariant components $\beta^{ab}$ of it are defined by the
relations
\begin{equation}
  \delta^a_b=\beta^{ac}\beta_{cb}=\beta_{bc}\beta^{ca}.
  \label{swe_topo_02_04.eq}
\end{equation}

The area element of the surface is defined by
\begin{equation}
  {\rm d}\sigma (y)=\beta (y){\rm d}y^1{\rm d}y^2,
  \label{swe_topo_02_05.eq}
\end{equation}
where
\begin{equation}
  \beta=\sqrt{\varepsilon^{ab}\beta_{a1}\beta_{b2}},
  \label{swe_topo_02_06.eq}
\end{equation}
with $\varepsilon^{ab}$ being the Levi-Civita symbol.

Note that
$$ ||\boldsymbol{\cal N}||=\beta. $$

\noindent {\bf The curvature tensor $\boldsymbol{\kappa}$}.
The curvature tensor $\kappa$ and the affine connection
$\gamma $ can be defined by the Gauss-Wiengarten equations
\begin{equation}
  \begin{array}{ll}
    \displaystyle 
    \frac{\partial \boldsymbol{\tau}_a}{\partial y^b}=
    \gamma^{c}_{a\,b}\boldsymbol{\tau}_c+\kappa_{a\,b}\boldsymbol{\nu},& ({\rm Gauss})\\
    \displaystyle 
    \frac{\partial \boldsymbol{\nu}}{\partial y^a}
    =-\kappa_a^b\boldsymbol{\tau}_b.& (Wiengarten)
  \end{array}
	\label{swe_topo_02_07.eq}
\end{equation}

\subsection{Surface Based Curvilinear Coordinate System}

A surface ${\cal S}$ based coordinate system in the space
$\mathbb{E}^3$ is introduced as follows.  Given a
parameterization (\ref{swe_topo_01_00.eq}) of the surface, one
defines the applications
\begin{equation}
  x^i=b^i(y^1,y^2)+y^3\nu^i,
  \quad (y^1,y^2)\in\widetilde{D}\subset \mathbb{R}^2,\quad
  y^3\in \widetilde{I}\in\mathbb{R},
	\label{swe_topo_03_1.eq}
\end{equation}
where $\widetilde{I}$ is an open neighborhood of zero.
Assume that (\ref{swe_topo_03_1.eq}) defines a coordinate
transformation from $\widetilde{D}\times\widetilde{I}$ to a
space neighborhood $\Omega$ of the surface ${\cal S}$.  The
surface ${\cal S}$ in the new coordinate system is given by
$y^3=0$.  Furthermore, we have:

$\bullet$ the tangent vectors  to the coordinate lines
\begin{equation}
  \boldsymbol{e}_I= 
  \frac{\partial \boldsymbol{x}}{\partial y^I}\Longrightarrow
  \left\{
  \begin{array}{l}
    \boldsymbol{e}_a=q^b_a\boldsymbol{\tau}_b, 
    \quad q^b_a:=\delta^b_a-y^3\kappa^b_a,\quad a=\overline{1,2}\\
    \boldsymbol{e}_3=\boldsymbol{\nu}
  \end{array}
  \right. ;
  \label{swe_topo_04.eq}
\end{equation}

$\bullet$ the coefficients of the metric tensor
\begin{equation}
  g_{IJ}=\delta_{ij}e_I^ie_J^j \Longrightarrow
  \left\{
  \begin{array}{ll}
    g_{ab}=q^c_aq^d_b\beta_{cd},&g_{a3}=0,\\
    g_{3a}=0,&g_{33}=1,
  \end{array}
  \right.
  \label{metric_01.eq}
\end{equation}
with
\begin{equation}
  \sqrt{{\rm det} \boldsymbol{g}}= 
\beta \Delta,\quad  \Delta:= 1-2y^3K_M+(y^3)^2K_G,
  \label{metric_01_second.eq}
\end{equation}
where $K_M=1/2\kappa^a_a$ and
$K_G=\epsilon_{a,b}\kappa^a_1\kappa^b_2$ are the mean
curvature and the Gauss curvature of the surface,
respectively;

$\bullet$ the affine connection
\begin{equation}
  \frac{\partial \boldsymbol{e}_I}{\partial y^J}
  =\Gamma^L_{IJ}\boldsymbol{e}_L\Longrightarrow
  \left\{
  \begin{array}{ll}
    \Gamma^{c}_{ab}=\left(\gamma^{d}_{ab}-y^3\left(
    \partial_a\kappa^d_b+\kappa^f_b\gamma^d_{af}
    \right)\right)Q^{c}_{d},&\Gamma^{c}_{a3}=-\kappa^e_aQ^c_e,\\
    \Gamma^3_{ab}=(\delta_a^{c}-
    y^3\kappa^{c}_a)\kappa_{cb},&\Gamma^3_{a3}=0,
  \end{array}
  \right.
  \label{metric_02.eq}
\end{equation}
where $Q$ is defined by
\begin{equation}
  \boldsymbol{\tau}_a=Q^b_a\boldsymbol{e}_b\Longrightarrow
  \left\{
  \begin{array}{ll}
    Q^{1}_1=\displaystyle\frac{1-y^3\kappa^2_2}{\Delta(y)},
    &Q^{2}_1=\displaystyle\frac{y^3\kappa^2_1}{\Delta(y)},\\
    Q^{1}_2=\displaystyle\frac{y^3\kappa^1_2}{\Delta(y)},
    &Q^{2}_2=\displaystyle\frac{1-y^3\kappa^1_1}{\Delta(y)}.
  \end{array}
  \right.
  \label{metric_03.eq}
\end{equation}

\noindent {\bf Obs.} For any $y^3\in I$, the tangent vectors
$\boldsymbol{e}_a$, $a=\overline{1,2}$ belong to the tangent
plane at the surface $y^3={\rm const}$ and they are
orthogonal to the normal
$\boldsymbol{e}_3=\boldsymbol{\nu}$.  In the new coordinate
system, the volume element is
$\vartheta(y){\rm d}y^1 {\rm d}y^2 {\rm d}y^3$, where
\begin{equation}
  \vartheta(y)=\epsilon_{i\,j\,k}e^i_1e^j_2e^k_3=
  \sqrt{{\rm det} \boldsymbol{g}} =
  \left(1-2y^3K_M+(y^3)^2K_G\right) \beta.
  \label{metric_04.eq}
\end{equation}

\subsection {Integrals of vectors and second order tensors}
Let $V$ be a domain in $\mathbb{E}^3$ defined by
$$
\boldsymbol{x}=\boldsymbol{b}(y^1,y^2)+y^3\boldsymbol{\nu}, \quad (y^1,y^2)\in
D ,\quad u(y^1,y^2)<y^3<w(y^1,y^2)
$$
where $D$ is a open closed domain with boundary
$\partial D$, $u(y^1,y^2)$ and $w(y^1,y^2)$ are two
functions that define some surfaces in $\mathbb{E}^3$.  We
are interested in calculating the flux of vectors or tensors
through the boundary of $V$, to evaluate integral of vectors
in $V$ or to calculate integrals of vectors on surfaces.  In
$\mathbb{E}^3$, such integrals define global quantities of
the same type with the integrands: scalars define scalars,
vectors define vectors and second order tensors define
second order tensors.  If one uses curvilinear coordinates,
such invariant properties are lost for vectors and tensors.

Let ${\cal S}$ and $V$ be a surface and a domain in
$\mathbb{E}^3$, respectively.  Define the flux of
$\boldsymbol{f}$ and $\boldsymbol{\Phi}$ through a surface
by
\begin{equation*}
  \begin{split}
    {\cal F}_{\boldsymbol{f}}(S)&:=\int\limits_{S}f^i\,n_i{\rm d}\sigma,\\
    {\cal F}^i_{\boldsymbol{\Phi}}(S)&:=\int\limits_{S}\Phi^{ij}\,n_j{\rm d}\sigma,
  \end{split}
\end{equation*}
where $\boldsymbol{n}$ stands for outward oriented unitary
normal to the surface.

Define by components the integral of a vector field
$\boldsymbol{f}$ on $V$ 
\begin{equation*}
    {\cal I}^j_{\boldsymbol{f}}(V):=\int\limits_{V}f^j{\rm d}x
\end{equation*}
and the integral on the surface $S$
\begin{equation*}
    {\cal I}^j_{\boldsymbol{f}}(S):=\int\limits_{S}f^j{\rm d}\sigma.
\end{equation*}

Let ${\cal S}_r$ be the surface defined by some function
$r(y^1,y^2)$
$$
\boldsymbol{x}=\boldsymbol{b}(y^1,y^2)+r(y^1,y^2)\boldsymbol{\nu},\quad (y^1,y^2)\in D.
$$
One denotes the ``vertical'' boundary of $V$ by
\begin{equation}
  \begin{split}
   \Sigma=\left\{\right .&\boldsymbol{x}\in \mathbb{E}^3|
    \boldsymbol{x}=\boldsymbol{b}(y^1(s),y^2(s))+
    y^3\boldsymbol{\nu}(y^1(s),y^2(s)),\\
    &s\in (0,L),\;\left.  u(y^1(s),y^2(s))<y^3<w(y^1(s),y^2(s))\right\}
  \end{split}
  \label{swe_topo_08.eq}
\end{equation}
where $\left(y^1(s),y^2(s)\right)$, $s\in (0,L)$ is a
parameterization of $\partial D$.

Let $\boldsymbol{f}$ and $\boldsymbol{\Phi}$ be a vector
field and a second order tensor field in $\mathbb{E}^3$,
respectively.  Using the law of transformation of the
coordinate system of a tensor field under coordinate
transformation, one can write
$$ 
f^i=f^I{e}^i_I, \quad \Phi^{ij}={e}^i_I {e}^j_J \Phi^{IJ}.
$$

\noindent
Next lemma refers to various integrals.

\begin{lemma}
  Let $\boldsymbol{f}$ and $\boldsymbol{\Phi}$ be some
  smooth fields on a domain $\Omega \subset \mathbb{E}^3$.
  Let $S_r$, $V$ and $\Sigma$ be a surface, domain and
  portion of $\partial V$, respectively, as previously
  defined.  Then:
  \begin{equation}
    \begin{split}
      {\cal I}^i_f(V)=
      & \displaystyle\iint\limits_{D}\left( 
      \tau^i_a\displaystyle\int\limits_{u}^{w}q^a_bf^b\vartheta{\rm d}y^3
      +\nu^i \int\limits_{u}^{w}f^3 \vartheta{\rm d}y^3
      \right){\rm d}y^1{\rm d}y^2,\\
      {\cal F}_f(S_r)=
      & \displaystyle\iint\limits_{D}\left.\vartheta  (y)
      \left(f^3-f^a\frac{\partial r}{\partial y^a}\right)\right|_{y^3=r}{\rm d}y^1{\rm d}y^2,\\
      {\cal F}_f(\Sigma)=
      & \displaystyle\iint\limits_{D}\frac{\partial r}{\partial y^a}\int\limits_{u}^{w}
      \vartheta f^a{\rm d}y^3{\rm d}y^1{\rm d}y^2,\\
      {\cal F}^i_{\Phi}(S_r)=
      & \displaystyle\iint\limits_{D}\left[\left( \tau^i_c\,q^c_b\left(\Phi^{b3}-
      \frac{\partial r}{\partial y^a}\Phi^{ba}\right) \right .\right .\\
      +&\left. \left.\left.\nu^i\left(\Phi^{33}-
      \frac{\partial r}{\partial y^a}
      \Phi^{3a}\right)\right)\vartheta(y)\right]\right|_{y^3=r}{\rm d}y^1{\rm d}y^2,\\
      {\cal F}^i_{\Phi}(\Sigma)=
      & \displaystyle\iint\limits_{D} \tau^i_c\left(
      \frac{\partial }{\partial y^a}\int\limits_{u}^{w}q^c_b\vartheta(y)\Phi^{ba}{\rm d}y^3\right .\\
      +&\left .\gamma_{ae}^c\int\limits_{u}^{w}q^e_b\vartheta(y)\Phi^{ba}{\rm d}y^3- 
      \kappa^c_a\int\limits_{u}^{w}\vartheta(y)\Phi^{3a}{\rm d}y^3\right){\rm d}y^1{\rm d}y^2\\
      +& \displaystyle\iint\limits_{D} \nu^i
      \left( \kappa_{ca}\int\limits_{u}^{w}q^c_b\vartheta(y)\Phi^{ba}{\rm d}y^3 \right .\\
      +&\left .\frac{\partial }{\partial y^a}\int\limits_{u}^{w}\vartheta(y)\Phi^{3a}{\rm d}y^3\right)
      {\rm d}y^1{\rm d}y^2.
    \end{split}
    \label{swe_lemma_01.eq}
  \end{equation}
\end{lemma}

\noindent {\it Proof}. Let $(y^1(s),y^2(s))$, $s\in (0,L)$
be a parameterization of the boundary $\partial D$.  On
$\Sigma$, the tangent directions are given
by 
\begin{equation*}
  \begin{array}{l}
    \boldsymbol{t}_s=\boldsymbol{e}_a w^a,\\
    \boldsymbol{e}_3=\boldsymbol{\nu},
  \end{array}
\end{equation*}
where $w^a=\displaystyle\frac{{\rm d}y^a}{{\rm d}s}$ and the
outward normal direction is given by
$$
N_i:=\epsilon_{jki}{e}_3^jt_s^k=\epsilon_{jki}\nu^j {e}_a^kw^a.
$$ 
Thus, one can evaluate the flux as
$$
{\cal F}_f(\Sigma) :=\int_{\Sigma}f^in_i{\rm d}\sigma=
 \int\limits_0^L\int
\limits_{\widetilde{u}(s)}^{\widetilde{w}(s)}
f^iN_i{\rm d}y^3{\rm d}s,
$$
with $\widetilde{w}(s)=w(y^1(s), y^2(s))$,
$\widetilde{u}(s)=u(y^1(s), y^2(s))$.  Then, one writes
$\boldsymbol{f}$ in the local basis
$\left\{\boldsymbol{e}_1, \boldsymbol{e}_2, \boldsymbol{e}_3\right\}$
and obtains
$$
f^iN_i=(f^b {e}_b^i+f^{3}\nu^i)N_i=\epsilon_{jki}\nu^j {e}_a^k
{e}_b^iw^af^b=\vartheta(y)\epsilon_{ab}w^af^b
$$
and
$$
{\cal F}_f(\Sigma) =
\int\limits_0^L\int\limits_{\widetilde{u}(s)}
^{\widetilde{w}(s)}\vartheta(y)\epsilon_{ab}w^af^b{\rm d}y^3{\rm d}s=
\int\limits_0^L\epsilon_{ab}w^a
\int\limits_{\widetilde{u}(s)}^{\widetilde{w}(s)}
\vartheta(y)f^b{\rm d}y^3{\rm d}s.
$$

Observe that
$\epsilon_{ab}w^a=\epsilon_{ab}\displaystyle \frac{\partial y^a}{\partial s}$
is the normal direction to the boundary $\partial D$ and use
the flux-divergence theorem and to obtain
\begin{equation}
  {\cal F}_f(\Sigma)=\iint\limits_{D}\frac{\partial }{\partial y^a}
  \int\limits_{u(y^1,y^2)}^{w(y^1,y^2)}\vartheta(y)f^a{\rm d}y^3{\rm d}y^1{\rm d}y^2.
  \label{swe_topo_09.eq} 
\end{equation}
On $S_r$, one has the tangent vectors
\begin{equation}
  \boldsymbol{\zeta}_a=\frac{\partial \boldsymbol{x}}{\partial y^a}
  =\boldsymbol{e}_a+\frac{\partial r}{\partial y^a}\boldsymbol{\nu}
  \label{swe_surface_01.eq}
\end{equation}
and normal direction
\begin{equation}
  N_i=\epsilon_{jki}\left ({e}_1^j+\frac{\partial r}{\partial y^1}\nu^j\right )
\left ({e}_2^k+ \frac{\partial r}{\partial y^2}\nu^k\right ).
  \label{swe_surface_02.eq}
\end{equation}
Then, we obtain
$$
f^iN_i=\vartheta(y)\left (f^3-
\displaystyle\frac{\partial r}{\partial y^a}f^a\right ).
$$
Consequently,
\begin{equation}
  {\cal F}_f(S_r)=\iint\limits_{D}
  \left.\vartheta(y)\left (f^3-
  \displaystyle\frac{\partial r}{\partial y^a}
f^a\right )\right|_{y^3=r}{\rm d}y^1{\rm d}y^2.
  \label{swe_topo_10.eq} 
\end{equation}
Consider now a second order tensor $\Phi$.  The coordinate
transformation (\ref{swe_topo_03_1.eq}) implies that the
contravariant components of the tensor in the two coordinate
system are related by
$$
\Phi^{ij}={e}^i_I {e}^j_J\Phi^{IJ}.
$$
The main difficulty in this case is that the vectors of the
basis depend on the variables $(y^1,y^2,y^3)$ and there is
no sense to find the components of the global vector
quantity ${\cal F}_{\Phi}$ in the new system of coordinates.  We
proceed to find the Cartesian components of
${\cal F}_{\Phi}$, but calculated as functions of the
contravariant components $\Phi^{IJ}$.

On the surface $\Sigma$, one has
$$
\Phi^{ij}N_j={e}^i_I {e}^j_J {\Phi}^{IJ}N_j=
\vartheta(y)\epsilon_{ab}w^a {e}^i_I\Phi^{Ib}
$$ 
and the flux is given by
$$
{\cal F}^i_{\Phi}(\Sigma)=
\iint\limits_{D} \displaystyle\frac{\partial }{\partial y^a}
\int\limits_{u}^{w} \vartheta(y){e}^i_I\Phi^{Ia} {\rm d}y^3{\rm d}y^1{\rm d}y^2.
$$ 
Using the relations (\ref{swe_topo_04.eq}) we get
$$
{\cal F}^i_{\Phi}(\Sigma)=\iint\limits_{D} \displaystyle\frac{\partial }{\partial y^a}\left( \tau^i_c\int\limits_{u}^{w}q^c_b\vartheta(y)\Phi^{ba}{\rm d}y^3 +\nu^i\int\limits_{u}^{w}\vartheta(y)\Phi^{3a}{\rm d}y^3 \right){\rm d}y^1{\rm d}y^2.
$$
Applying Weigartern formula, we can write
\begin{equation*}
  \begin{split}
    {\cal F}^i_{\Phi}(\Sigma)=& 
    \displaystyle\iint\limits_{D}\left( \tau^i_c\frac{\partial }{\partial y^a}
    \int\limits_{u}^{w}q^c_b\vartheta(y)\Phi^{ba}{\rm d}y^3 
    +\nu^i \displaystyle\frac{\partial }{\partial y^a}
    \int\limits_{u}^{w}\vartheta(y)\Phi^{3a}{\rm d}y^3
    \right){\rm d}y^1{\rm d}y^2\\
    &+\displaystyle\iint\limits_{D}
    \tau^i_c\left(
    \gamma_{ae}^c\int\limits_{u}^{w}q^e_b\vartheta(y)\Phi^{ba}{\rm d}y^3
    -\kappa^c_a\int\limits_{u}^{w}\vartheta(y)\Phi^{3a}{\rm d}y^3\right){\rm d}y^1{\rm d}y^2\\
    &+ \displaystyle\iint\limits_{D} \nu^i\kappa_{ea} \int\limits_{u}^{w}q^e_b\vartheta(y)\Phi^{ba}{\rm d}y^3 {\rm d}y^1{\rm d}y^2.
  \end{split}
\end{equation*}
Regrouping the terms, we obtain the result for
${\cal F}^i_{\Phi}(\Sigma)$.

\begin{lemma}
  Consider that the stress tensor of the fluid has the
  following form
$$
t^{ij}=-p\delta^{ij}+\tau^{ij}
$$
and set
$$
{\cal F}^i_{\rm stress}(S_r)=\iint\limits_{S_r}t^{ij}n_j{\rm d}\sigma.
$$
Then 
\begin{equation}
  \begin{split}
    {\cal F}^i_{\rm stress}(S_r)=
    &\displaystyle\iint\limits_{D}\left[ 
    \tau^i_{d}q^{d}_a\left((p-\widetilde{\tau}^{33})
    g^{ab}\frac{\partial r}{\partial y^b}\right .\right .\\
    + &\left .\left . \left .\widetilde{\tau}^{a3}\sqrt{1+g^{bc}
    \frac{\partial r}{\partial y^{b}}
    \frac{\partial r}{\partial y^{c}}}\right) 
    \vartheta(y)\right]\right|_{y^3=r(y^1,y^2)}{\rm d}y^1{\rm d}y^2\\
    +&\displaystyle\iint\limits_{D}
    \left [ \nu^{i}
    \left (-p+\widetilde{\tau}^{33} +\frac{\partial r}{\partial y^a}
    \widetilde{\tau}^{a3}
    \right .
    \right .\\
    \cdot&
    \left . 
    \left .
    \left .
    \sqrt{1+g^{bc}\frac{\partial r}{\partial y^{b}}
    \frac{\partial r}{\partial y^{c}}}
    \right) \vartheta(y)
    \right ]
    \right |_{y^3=r(y^1,y^2)}{\rm d}y^1{\rm d}y^2.
  \end{split}
  \label{swe_lemma_02.eq}
\end{equation}
\end{lemma}

In this lemma, $\widetilde{\tau}^{IJ}$ denotes the
contravariant components of the viscous stress tensor in the
frame given by the tangent vectors to the surface
$y^3=r(y^1,y^2)$ and the unit normal to the tangent plan
(which points to the same direction as the unit normal
$\boldsymbol{\nu}$ to the support surface).

\noindent {\it Proof.} Let $r(y^1,y^2)$ be a parameterization
of the surface $S_r$ and let $\boldsymbol{\zeta}_1$,
$\boldsymbol{\zeta}_2$ and $\boldsymbol{n}$ be the tangent
vectors and the unit normal given by
(\ref{swe_surface_01.eq}) and (\ref{swe_surface_02.eq}),
respectively.  One can write
\begin{equation}
  t^{ij}n_j= -pn^i+\tau^{ij}n_j=-pn^i+\widetilde{\tau}^{a3}\zeta^i_a+\widetilde{\tau}^{33}n^i.
  \label{swe_stress_01.eq}
\end{equation}
Using the basis $\{\boldsymbol{e}_I\}$, the unit normal has
the form
\begin{equation*}
  \begin{array}{ll}
    \boldsymbol{n}=n^a\boldsymbol{e}_a+n^3\boldsymbol{\nu},&\displaystyle
    n^a=-g^{ab}\frac{\partial r}
    {\partial y^{b}}\frac{\vartheta(y)}{||\boldsymbol N||},
    \quad n^3=\frac{\vartheta(y)}{||\boldsymbol N||},\\
    &\displaystyle||\boldsymbol N||=\vartheta(y)\sqrt{1+g^{ab}
    \frac{\partial r}{\partial y^a}
    \frac{\partial r}{\partial y^b}}, \quad y^3=r(y^1,y^2)
  \end{array}
\end{equation*}
and the tangent vectors are expressed by
$$
\displaystyle
\boldsymbol{\zeta}_a=\boldsymbol{e}_a+
\frac{\partial r}{\partial y^a}
\boldsymbol{nu}.
$$
Since the area element is given by
$$
{\rm d}\sigma=||\boldsymbol N||{\rm d}y^1{\rm d}y^2,
$$
then, we immediately obtain the conclusion of this lemma.

\section*{Acknowledgment}
This work was partially supported by grants of the Ministry
of Research and Innovation, CCCDI-UEFISCDI, project number
PN-III-P1-1.2-PCCDI-2017-0721/34PCCDI/2018, and project
50/2012 ASPABIR.

\end{document}